\documentclass[11pt]{article}
\usepackage{graphicx}
\usepackage{amssymb}
\def\mdmatm{\Delta m^2_{31}}
\def\dmatm{$\mdmatm$}
\def\mdmsol{\Delta m^2_{21}}

\def\lsim{\mathrel{\raise.3ex\hbox{$<$\kern-.75em\lower1ex\hbox{$\sim$}}}}
\def\gsim{\mathrel{\raise.3ex\hbox{$>$\kern-.75em\lower1ex\hbox{$\sim$}}}}

\begin{document}

\title{Precision physics with a wide band super neutrino beam }
\author{V. Barger$^1$, M. Dierckxsens$^2$, M. Diwan$^2$,
P. Huber$^1$, C. Lewis$^3$, D. Marfatia$^4$ and B. Viren$^2$\\[2ex]
\small\it $^1$Department of Physics, University of Wisconsin, Madison, WI
53706\\
\small\it $^2$Brookhaven National Laboratory, Upton, NY  11973\\
\small\it $^3$Department of Physics, Columbia University, New York, NY
10027\\
\small\it $^4$Department of Physics and Astronomy, University of Kansas, 
Lawrence, KS 66045
}

\date{}

\maketitle
%\author{V. Barger, Patrick Huber}
%\affiliation{\em Department of Physics, University of Wisconsin, Madison, WI 53706} 

%\author{Danny Marfatia}
%\affiliation{\em Department of Physics and Astronomy, University of Kansas, Lawrence, KS 66045} 

%\author{Mark Dierckxsens,  Milind V. Diwan, Brett Viren}
%\affiliation{\em Brookhaven National Laboratory, Upton, NY 11973} 

%\author{Christine Lewis}
%\affiliation{\em Department of Physics, Columbia University, New York, NY 10027} 

\begin{abstract}

We carry out a state-of-the-art assessment of long baseline 
neutrino oscillation 
experiments with wide band beams.
We describe the feasibility of an  experimental program 
%in neutrino physics  
using
existing high energy accelerator facilities, a new intense wide band neutrino 
beam (0--6 GeV)  and
a proposed large detector in a deep underground laboratory.
We find that a decade-long program 
 with 1~MW operation in the neutrino mode
and 2~MW operation in the antineutrino mode,
 a baseline as long as the distance between Fermilab and the 
Homestake mine (1300 km) or the
Henderson mine (1500 km), and
a  water Cherenkov detector with fiducial mass $\sim 300$ kT 
%(or a smaller liquid argon detector with better  performance)
 has optimum
 sensitivity to a nonzero $\theta_{13}$, the mass hierarchy and to
neutrino CP violation at the $3\sigma$~C.~L. for $\sin^2 2\theta_{13}>0.008$.
This program is capable of breaking the eight-fold degeneracy down to the
octant degeneracy without additional external input. 
%We find that the sensitivity has a broad optimum for baselines $\gsim 1000$km.
%The dependence of the sensitivity on the solar and atmospheric 
%oscillation parameters  and on 
% the  length of the baseline beyond $1000$ km
%  is weak, making this  a robust experimental program. 

\end{abstract}

\newpage
%\maketitle
%
%\baselineskip=17pt
%
\section{Introduction}

There is abundant evidence that neutrinos oscillate and hence have 
mass~\cite{Barger:2003qi}.
Atmospheric neutrinos oscillate with a mass-squared difference 
$|\Delta m^2_{31}| \sim 0.0025$ eV$^2$ and mixing angle 
$\theta_{23}\sim \pi/4$~\cite{superkatm} while the
corresponding parameters for solar neutrinos are 
$\mdmsol \sim 8\times 10^{-5}$ eV$^2$ 
and $\theta_{12} \sim \pi/6$~\cite{Eguchi:2002dm}. The 
unknowns are the angle $\theta_{13}$ coupling the 
atmospheric and solar sectors, the CP-violating phase $\delta_{\rm CP}$, and the sign 
of \dmatm, which fixes the hierarchy of neutrino masses.
All we presently know
about $\theta_{13}$ is that
$\sin^2 \theta_{13} < 0.05$ 
at the 2$\sigma$ C.~L.~\cite{Apollonio:1999ae}. 
   Measurements of the unknown parameters are the main goals of 
future long-baseline neutrino experiments. Complete and precise 
experimental knowledge of
 neutrino parameters is essential to test neutrino mass models. 
Moreover, a determination of the type of mass hierarchy is of 
 importance for the feasibility of observing neutrinoless double beta
% decay and determining the Majorana nature of the neutrino 
if the heaviest neutrino is $\cal{O}$(0.05) eV. 
An observation of this decay will
 provide much confidence in the Majorana nature of neutrinos.   

In the near future, experiments with low energy conventional neutrino beams 
will attempt to detect $\nu_\mu \to \nu_e$ transitions at the atmospheric
scale. Concurrently, reactor experiments will try
to determine or constrain $\theta_{13}$ by observing $\bar{\nu}_e$ 
disappearance at the atmospheric scale. However, the complementary 
information gathered from 
both types of experiments will be inadequate to determine the mass hierarchy
or to establish  CP-violation in the lepton sector.
In recognition of this fact, planning is underway for future facilities
that will address these outstanding issues.

Two types of neutrino beams are being considered for future long-baseline 
experiments: off-axis beams with narrow band energy 
spectra (as will be used in the Tokai-to-Kamioka (T2K) 
experiment at Japan Proton Accelerator
Proton Complex (J-PARC)~\cite{jhfsk}, and proposed for the NuMI Off-axis 
$\nu_e$ Appearance Experiment or NO$\nu$A~\cite{nova})  
and a wide band beam~\cite{wide-band,previous} 
for which the neutrino energy 
is inferred 
from quasielastic scattering events. 

Off-axis beams~\cite{e889} 
are advocated because their beam energy is narrow 
and depend to first order only on the kinematics of pion decay,
so that a counting experiment can be carried out at an 
oscillation maximum. Another positive feature of the narrow beam is the
reduction of  a high energy tail which could contribute background  to
reconstructed events at low energy via neutral-current feed down. 
%Moreover, 
%off-axis beams have a reduced
%$\nu_e$ background. 
The features that make off-axis beams attractive also
lead to their limitations. The very nature of a counting experiment precludes
the possibility of using spectral energy information. 
Consequently, it is virtually
impossible to resolve the eight-fold parameter degeneracy~\cite{bmw} 
(which arises from three independent two-fold degeneracies 
$(\delta_{\rm CP},\theta_{13})$, 
sgn$(\Delta m^2_{31})$  and 
$(\theta_{23},\pi/2-\theta_{23})$~\cite{bmw,b-c}), unless multiple 
measurements with separate detectors 
are made at different off-axis angles  and/or baselines.
The capabilities and optimization
of experiments with off-axis beams have been studied 
extensively~\cite{off-axis}.

Wide band beams possess the advantages of a higher on-axis flux and a broad
energy spectrum. In an idealized sense, one may think of them as enabling
many off-axis experiments in a single experiment. Thus, parameter
degeneracies may be resolved in a single wide band beam experiment. 
%The disadvantages arise because wide band beams typically have higher energy,
%making it necessary to have a longer baseline to probe the
%first oscillation maximum. This results in a reduced flux and the
%requirement of very large detectors. 
The disadvantage of a wide band beam is the need to determine the neutrino 
energy with good resolution and eliminate background 
from high energy tails of the spectrum. Therefore,
detectors with good energy resolution and neutral-current rejection are needed
to use the broad spectrum fruitfully and to reject the feed down
from the high-energy
tail efficiently. 
%In sum, the demands on the detector are quite stringent.

The possibility of working with large detectors may not be too far into
the future~\cite{3m,uno,icarus}. 
Two concrete proposals
are under consideration  for a Deep Underground Science and
Engineering Laboratory (DUSEL)~\cite{nusel} in the U.S., 
one at the Homestake mine in South Dakota, 
the other at the Henderson mine in Colorado. DUSEL will be designed to 
accomodate 
Mton class multipurpose  detectors. 
The physics program of these detectors will include nucleon decay,
supernova neutrino detection, as well as long-baseline neutrino 
oscillation physics with an accelerator beam.

%While preliminary 
%studies of experiments with wide band beams and large detectors 
%have been made, 
%a comprehensive analysis is lacking. 
%Our purpose is to fill this gap.
%In previous work, 
%the BNL collaboration has analyzed 
The prospects of
a very long-baseline experiment  
 consisting of a wide band super neutrino 
beam and a large water Cherenkov detector were analysed in 
Refs.~\cite{previous,hql04}, where, in the spirit of a first analysis, 
correlations
and degeneracies were only partially accounted for.
Preliminary work had suggested that 
sufficient background suppression could be achieved so as to extract the
quasielastic events with sufficient purity. 
Since then, further detailed  assessments  with a full simulation of a
water Cherenkov detector have been performed~\cite{chiaki}. 
The main result of that work is that neutral current backgrounds 
can be sufficiently suppressed while maintaining good 
signal efficiency with new software techniques.  The remaining
signal events are dominated by quasielastics, but have significant 
 contamination of other $\nu_e$ induced charged current events. Nevertheless, 
 neutrino energy resolution in the range of $\sim 10\%$ can be maintained.  
This work will be reported elsewhere in detail. 

In this paper, we  perform
 a thorough state-of-the-art analysis
with a realistic treatment
of systematic errors, correlations and degeneracies so as to define
an optimum program using a wide band beam.
We consider accelerator and detector requirements for
a feasible wide band experiment in the next two sections.
 In Section 4 we discuss our analysis methodology and present results
of the analysis in Section 5. We summarize in Section 6.

\section{Accelerator Requirements}
A number of recent  studies have  examined possible intensities 
and neutrino spectra from U.S. proton 
accelerators~\cite{foster, fnalmarch,agsup}.  Here we summarize the  
understanding of the intensity versus proton energy available  in the 
U.S.  currently, in the near future, and with  upgrades.  

High energy multi-MW proton beams are under consideration at
FNAL.  An examination is underway to increase the total power 
from the 120~GeV  Main Injector (MI) complex after the Tevatron program 
ends~\cite{mcginnis}. In this scheme protons from the 8~GeV booster, 
operating at 15~Hz, will be stored in the antiproton accumulator 
(which becomes available after the
shutdown of the Tevatron program)  while the MI 
completes its acceleration cycle. 
 Combining the techniques of
momentum stacking
using the antiproton accumulator and slip-stacking using the recycler 
will raise the total intensity in the MI to
$\gsim 1$~MW at 120~GeV. 
In the ideal case,  the length of the acceleration cycle is proportional
to the proton energy, making the average beam power proportional to
the final proton energy.
However, there are fixed time intervals in the beginning and the
end of the acceleration cycle for stable operation. These become
important at low
energies and reduce the performance below the ideal. Current
projections suggest that
$\sim 0.5$ MW operation between $40-60$~GeV and $\gsim 1$~MW operation at
120~GeV is possible.

Ambitious plans at FNAL call for a 8~GeV
super-conducting LINAC that can provide $1.5\times 10^{14}$ $H^-$ ions
at 10~Hz corresponding to 2~MW of total beam power~\cite{foster}. Some of the 
8~GeV ions could be injected into the MI to provide high
proton beam power, 1 to 2 MW, at  energy between 40 and 120 GeV; {\it e.g.},
40~GeV at $\sim 2$~Hz or 120~GeV at $\sim0.67$~Hz. 
Such a plan allows for
flexibility in the choice of proton energy for neutrino production. 

The BNL Alternating 
Gradient Synchrotron 
(AGS)  operating at 28~GeV
currently can provide about 1/6~MW of beam power. This corresponds to 
an intensity of about $7\times 10^{13}$ protons in a 2.5 microsecond
pulse every 2~seconds. The AGS complex can be upgraded 
 to provide a total  proton beam 
power of 1~MW~\cite{agsup}. 
The main components of the accelerator upgrade at BNL are 
a new 1.2~GeV Superconducting LINAC to provide protons to the 
existing AGS, and new magnet power supplies to increase the ramp rate of the 
 AGS magnetic field from about 0.5 Hz to  2.5 Hz.  
For 1 MW operation the protons from the accelerator will be delivered 
in pulses of $9\times 10^{13}$ protons at 2.5 Hz.
It has been determined that 2 MW operation of the AGS is also possible by 
further upgrading the synchrotron to 5 Hz repetition rate and with 
further modifications to the LINAC and the RF systems. 
A new neutrino beam could be built with conventional horn focussed 
technology and a 200 m long pion decay tunnel placed on  the slope 
of a specially 
built hill. Such a beam could be aimed at a detector $\sim 2500$~km away. 
%The spectra,
%and the physics sensitivity  for such a beam coupled to a large 
%detector have been considered in Ref.~\cite{previous, hql04}.  

To observe multiple 
oscillation nodes in the $\nu_\mu$ disappearance channel,
it is necessary to have a wide band beam with
energies from 0 to 6 GeV. Protons with energy 
above $\sim20$ GeV are needed to provide 
such a flux, which is clearly possible at both FNAL and BNL. 
The spectrum that is used in this paper was obtained 
using 28 GeV protons and a 
200 meter long decay tunnel. 
For details of this spectrum see Refs.~\cite{hql04,specwble}. 
Recent calculations have shown that similar, but more intense 
spectra per unit  beam power can be obtained using 40 or 60 GeV 
protons from FNAL, 
where the meson decay tunnel could be made longer than 
200 meters~\cite{mbishai}. 

In our analysis we assume that the spectrum from either the FNAL or the
BNL beam will be the same with a total average beam power of 1 MW.    
This allows us to make a proper comparison of
the physics issues regarding the baselines.
We comment on the additional flexibility obtained by the ability to 
change the proton energy at FNAL as well as the longer decay tunnel.  
We also comment on the impact on the sensitivity with
0.5 MW of average power from FNAL. It should be noted 
that 0.5 MW operation from 
FNAL does not appear to require a major upgrade.

If a large detector~\cite{3m,uno,icarus}
 is located at
Homestake (HS) or Henderson (HD), the beam from FNAL (BNL) 
will have to traverse a distance of
1290 km or 1495 km (2540 km or 2770 km), respectively. 
 At FNAL the
inclination will be about $5.8^\circ$ (HS) or $6.7^\circ$ (HD). 
The existing experience at FNAL, from
building the NuMI beam,  
could be extended to build  a new beam to HS or HD.
At BNL the beam would 
have to be built at
an inclination of about $11.4^\circ$ (HS) or  $12.4^\circ$ (HD). 
 Current design for such a beam
requires the construction of a hill with a height of about 
50~m~\cite{agsup}.  Such a hill will have the proton target at the top of
the hill and a 200 m long decay tunnel on the downslope.
 In either case, 
it is adequate to have a decay
tunnel  with length shorter (200 to 400 m) than  the NuMI tunnel 
(750 m) to obtain the needed  flux. Since the focus is on lower
energies, the beam intensity is maintained if the decay tunnel can be 
made about 4~m in diameter. 

Another  advantage of the wide tunnel is that 
the option of running with a narrow band beam using the off-axis
technique could be preserved.  
A 4~m diameter tunnel could permit
the rotation of the target and horn assembly so that a $1^\circ$ 
off-axis beam could be sent to the far detector.

\section{Detector Requirements}

%Important considerations for such a
%detector are the fiducial mass, energy threshold, energy resolution,
%muon/electron discrimination, pattern recognition capability, time
%resolution, depth of the location, and the cost. 
%Two types of
%detectors are under consideration for a large detector facility at 
%DUSEL: a water Cherenkov detector
%instrumented with photo-multiplier tubes and a liquid Argon based time
%projection chamber.

A  water Cherenkov detector with 300 kT fiducial mass 
can be built in the same manner as the 
SuperKamiokande detector (with 20 inch photo-multipliers placed on the
inside detector surface covering 
approximately 40\% of the total area)~\cite{sknim} by  
simply scaling it to larger size or by
building several detector modules~\cite{3m,uno}.  Such a detector
placed underground at DUSEL could have a low energy threshold 
($\lsim 10$~MeV), good energy resolution ($\sim 10\%$) for single 
particles, good
muon/electron separation ($\lsim1$\%), and time resolution ($\lsim$ few ns).
As noted earlier, it is important to obtain good
energy resolution when using a wide band beam. This can be achieved in a
water Cherenkov detector by separating quasielastic scattering events
with well identified leptons in the final state from the rest of the
charged-current (CC) events. The fraction of quasielastics in the total
CC rate with the spectrum used in this paper is about
23\% for the neutrino beam and 39\% for the antineutrino
beam.

Separation of quasielastic events from the CC
and neutral-current (NC) background is being used in the 
K2K experiment~\cite{k2k}.
For the program considered here an essential problem is to separate 
electron shower events from other NC events, especially events 
containing a single $\pi^0$ in the final state. The goal is to search for 
$\nu_e$ induced showering events in the 0.5 to 4 GeV range.  
Single $\pi^0$ particles with energies of 1, 2, 3 and 4~GeV 
decay to two photons 
with a minimum and most probable 
opening angle of 16, 8,  5, and 4 degrees, respectively. 
The probability of a decay with an opening angle of more than $20^\circ$
for 1, 2, 3 and 4 GeV $\pi^0$'s is 40\%, 8.2\%, 3.6\%, and 2.0\%, 
respectively.   
In a water Cherenkov detector the position where the $\pi^0$ photons 
convert cannot be measured with sufficient precision from the pattern of 
Cherenkov light which tends to 
be two overlapping showering rings.  At low $\pi^0$ energies the opening 
angle is sufficiently 
large compared to the Cherenkov angle ($42^\circ$) that single $\pi^0$'s 
can be separated 
quite effectively.  At energies greater than 2 GeV, however, the small 
angular separation between 
the two photons makes such separation difficult. 
 It is well known that resonant single pion
production in neutrino reactions has a rapidly falling cross section
as a function of momentum transfer, $q^2$, up to the
kinematically allowed value~\cite{adler}. 
This characteristic alone suppresses the
background by more than 2 orders of magnitude for $\pi^0$ (or shower)
energies above 2 GeV~\cite{bnl69395}.
Therefore a modest $\pi^0$ background suppression (by a factor of
$\sim 15$ below 2 GeV and $\sim 2$ above 2 GeV) makes
the $\pi^0$ background manageable level over the entire spectrum.
Such background suppression has recently been demonstrated using 
complete simulation and reconstruction  in Ref.~\cite{chiaki}.  
%Further
%work is needed to make this event reconstruction work at higher
%energies. The reconstruction algorithm could be enhanced by 
%the addition of ring imaging techniques to the detector\cite{ypsilantis}.

It has been argued that a liquid Argon time projection chamber (LARTPC) 
could be built with total mass approaching 100 kT~\cite{icarus}. 
A fine grained detector such 
as this has excellent resolution for separating tracks, making it possible 
to use a 
large fraction of the CC events
(rather than only the quasielastic events) 
to determine  the neutrino energy 
spectrum. A LARTPC also has much better particle identification 
capability. Consequently, a LARTPC with a 
total fiducial mass of $\sim100$ kT is expected to have similar 
performance as a 300 kT water Cherenkov detector. 
 
\begin{figure}
\centering\leavevmode
\includegraphics[angle=0,width=1\textwidth]{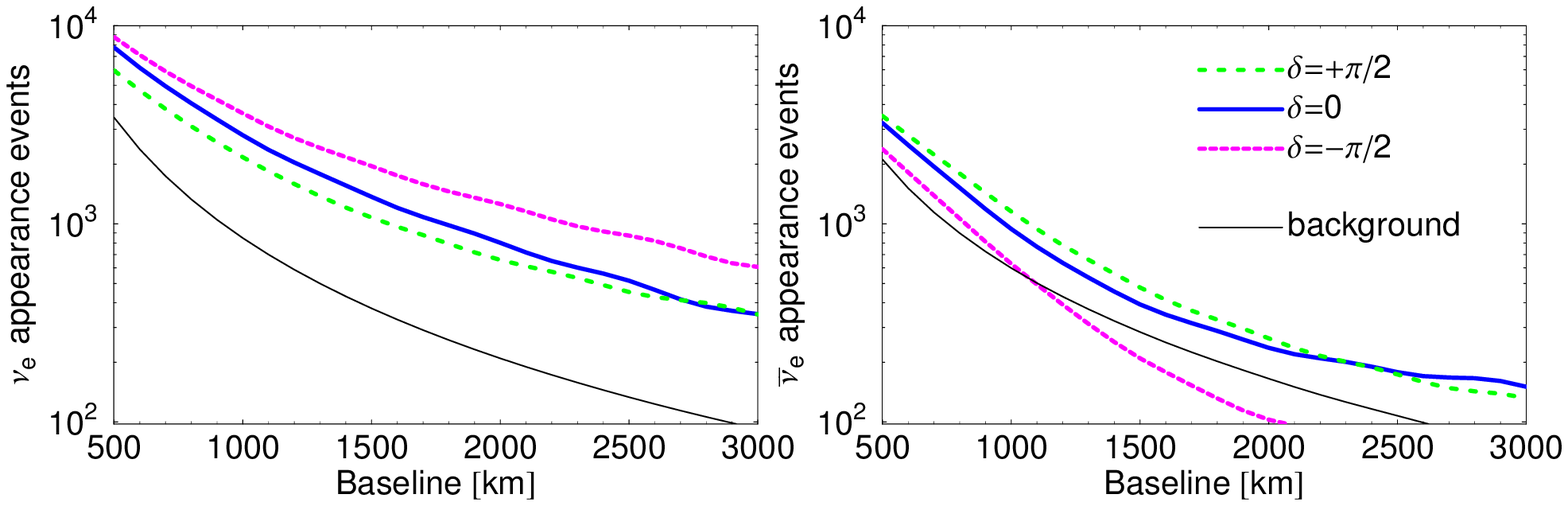}
 \caption{ 
  \label{rates} Event rates for neutrinos (left-hand panel) and
  antineutrinos (right-hand panel) as a function of baseline for
  $\sin^22\theta_{13}=0.1$ and a normal hierarchy. 
  The bold lines show the signal for 
  various choices of
  $\delta_{\rm CP}$. The thin line shows the total
  background, which includes background from beam contamination and neutral current 
events. }
\end{figure}

We assume a detector performance based on Ref.~\cite{chiaki}. 
For the physics sensitivities we 
assume 1 MW operation for 5 years (at $1.7\times 10^7$ sec/yr) 
in the neutrino mode, 
2 MW operation for 5 years  in the antineutrino mode and a 
detector fiducial mass of 300 kT.
 With the running times, the accelerator 
power level, and the detector mass fixed, we consider baselines
in the range of 500 to 3000 km. 
With 1 MW of a 28 GeV proton beam, a baseline of 1300 km, 
and a 300kT fiducial volume 
detector we calculate $\sim 230000$ muon charged
current and $\sim 77000$ neutral current events in 5 years of  
running in the neutrino mode in the absence of oscillations.  Under the
same running conditions in the antineutrino mode (with the horn
current reversed) we find a total of $\sim 74000$ antimuon charged
current and $\sim 27000$ neutral current events; approximately 20\% of the
event rate in the antineutrino beam will be due to wrong-sign
neutrino interactions. 
%For baselines from 500 to 3000 km,
% the total event rate will follow the $1/L^2$ law. 
For both neutrino and antineutrino running, approximately $\sim0.7$\% of the
CC rate will be from electron CC events
which form a background to the $\nu_\mu \to \nu_e$ search. 
As an example, the total number of 
electron neutrino appearance events and the expected background
for $\sin^2 2 \theta_{13}=0.1$ and a normal hierarchy
  as a function of baseline
are shown in Fig.~\ref{rates}.

The event rates for other beam configurations 
using 40 or 60 GeV proton beam have been calculated in Ref.~\cite{mbishai}.
For equal proton beam power, higher proton energy could result in 
as much as a $\sim50$\% increase in the total event rate with 
a concomitant increase in the background. A careful 
evaluation of the optimization of event rate versus background 
could reduce the running time 
by a significant factor. Preliminary calculations suggest that 
the physics sensitivity using other beam configurations will 
remain approximately the same as the calculations reported in this 
paper after accounting for the effects of spectral variations on the 
backgrounds and signal.

%\input{det.tex} 

%\input{anal.tex} 

%%%%%%%%%%%%%%%%%%%%%%%%%%%%%%%%%%%%%%%%%%%%%%%%%%%%%%%%%%%%%%%%%%%%%%%%%%%

\section{Analysis techniques}

We   simulate the expected event counts for the 
experimental program described above and 
compute
physics sensitivities using the GLoBES software
package~\cite{globes}. 
The description of the detector performance
follows from the results in Ref.~\cite{chiaki}. 
We used this detector performance to create large numbers 
of Monte Carlo events for the following set of event classes, 
which in turn, are used  
 to create migration matrices suitable for GLoBES:
\begin{itemize}
\item $\nu_e$ appearance signal 
\item $\nu_\mu$ disappearance signal 
\item $\nu_e$ beam events
\item NC background mainly consisting of 
single $\pi^0$ production. 
\item Other NC events. 
\item CC events with additional 
undetected particles. 
\end{itemize}
We have a set of six matrices for neutrinos and antineutrinos each.
The background of antineutrinos in the
neutrino beam is very small since it is suppressed both
in production and detection by the cross section ratio of neutrinos
to antineutrinos. For antineutrino running the opposite is true.
There the neutrino (wrong-sign) contamination 
 in the beam is nonnegligible due
to the enhancement by the cross section ratio. Therefore, we account for
this background via 6 additional matrices that
correspond to the ones above for treating the neutrinos in the
antineutrino beam. 
Our calculations show that the background due to tau neutrino interactions is
low and can be safely ignored because the spectrum we
 use has low energy. When combined with the high threshold (3.5~GeV)
for tau production as well as the low  $\nu_\tau$ CC 
cross section, the background is not significant. If higher
energy protons are used to create a spectrum with a high energy tail,
then this background must be included.
We have carefully checked that the GLoBES
calculation of event rates agrees with the direct result of the Monte
Carlo for the case of no oscillations.

Once event rates are computed, the next step is the calculation of a
$\chi^2$ function and the inclusion of systematic errors.  We use a
$\chi^2$ function for Poissonian processes as given, {\it e.g.},
in Ref.~\cite{ref1}. Systematic errors are implemented using a pull
approach as in Ref.~\cite{Huber:2002mx}. The systematics considered are a
normalization error for each type of signal, {\it i.e.}, $\nu_e$
appearance, $\bar\nu_e$ appearance, $\nu_\mu$ disappearance and
$\bar\nu_\mu$ disappearance of $1\%$ each, uncorrelated between the
four types of signal. For the sum of all backgrounds to each signal we
assume a $10\%$ uncertainty in the normalization, again uncorrelated
between the four types of signal.

As input or true values for the oscillation parameters and the Gaussian 
$1\sigma$ ranges (in anticipation of precision measurements from near future
experiments) we use:
\begin{eqnarray}
\label{eq:oscp}
\theta_{12}=0.55 \pm 10\%\,,&\quad& \Delta
m^2_{21}=(8.0\pm0.8)\cdot10^{-5}\,\mathrm{eV}^2\,,\nonumber\\
\theta_{23}=\pi/4 \pm 5\%\,,&\quad& \Delta m^2_{31}=(2.5\pm0.125)\cdot10^{-3}\,\mathrm{eV}^2\,.
\label{equation}
\end{eqnarray}
We include a $5\%$ error on the matter density. Our analysis
includes the correlations between all parameters and properly
accounts for possible degeneracies.

All calculations assume a normal mass hierarchy as input, whereas the 
fit always extends to the case of an inverted
  hierarchy. It is known that there are no qualitative
differences in the sensitivities if an inverted hierarchy is assumed as the
input; see {\it e.g.}, Ref.~\cite{Huber:2005jk}. 
In that case the matter enhancement  moves
from neutrinos to antineutrinos, but since we assume nearly symmetric
neutrino and antineutrino running, the results
are not affected significantly.

To allow a concise presentation of the physics results we define
performance indicators for the various measurements. For $\theta_{13}$
we choose the discovery of nonzero $\sin^22\theta_{13}$. Here a
nonzero value of $\sin^22\theta_{13}$ is chosen as the true value in
the simulation and a fit with $\sin^22\theta_{13}=0$ is performed.
The reach for discovering CP violation is computed by choosing a value for
$\delta_{\rm CP}$ as input and fitting it with the two CP conserving values of
$\delta_{\rm CP}=0,\pi$. All  other parameters (including $\theta_{13}$)
 are free  and the cases of
an inverted mass hierarchy and $\theta_{23}\rightarrow\pi/2-\theta_{23}$
are taken into account.
For the exclusion of an inverted mass hierarchy a
point in parameter space with normal mass hierarchy is chosen as the true
value and the solution with the smallest $\chi^2$ value with inverted
hierarchy has to be determined (global minimum  of the
$\chi^2$ function). The same applies to the resolution of the octant
of $\theta_{23}$. Data is generated with $\theta_{23}<\pi/4$ and the
global minimum of the $\chi^2$ function with  $\theta_{23}>\pi/4$ has
to be found.

The $\chi^2$ values obtained by the above procedure are
converted to confidence levels by using the $\chi^2$ distribution for
1 degree of freedom for each of the above cases, 
{\it i.e.}, $\sqrt{\Delta\chi^2}$ corresponds to the
significance in Gaussian standard deviations.

\section{Results}

\begin{figure}
%\vskip0.5in
\centering\leavevmode
\includegraphics[angle=0,width=0.5\textwidth]{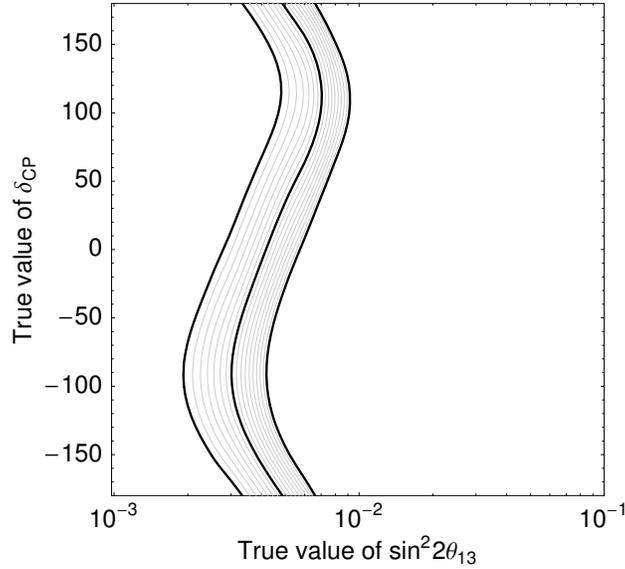}
 \caption{ 
   Discovery potential for $\sin^22\theta_{13}\neq 0$ at a baseline of
   $1300\,\mathrm{km}$. 
The bold iso-$\chi^2$ lines are
   $3,4,5\,\sigma$ (from left to right) and the light lines show an
   increase of $\chi^2$ by $1$.
 For all points to the right of the
   rightmost bold line, a nonzero value of $\sin^22\theta_{13}$ can be
   established with at least $5\sigma$ significance.
  \label{th13disc1300} }
\end{figure}

\begin{figure}
\centering\leavevmode
\includegraphics[angle=-90,width=\textwidth]{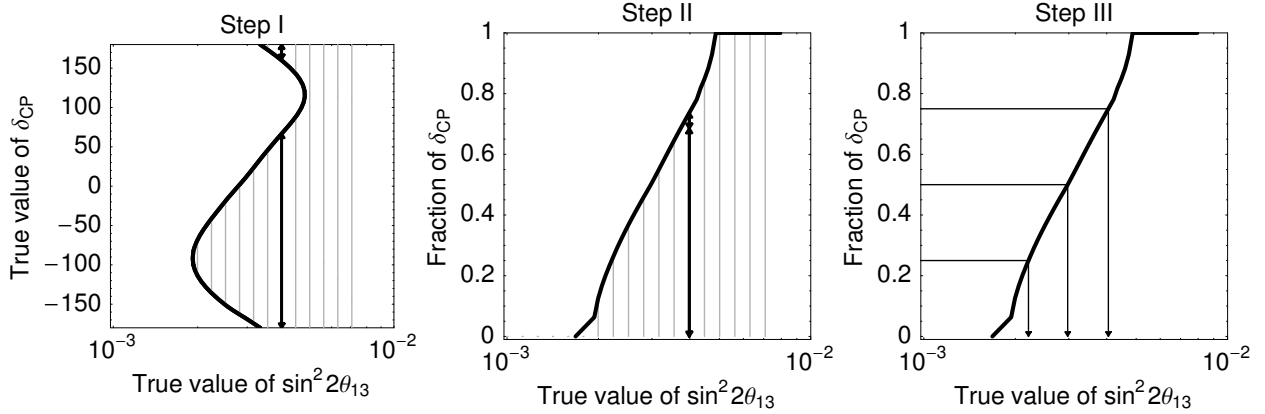}
 \caption{
   Schematic definition of the CP fraction. First, for each value of
   $\sin^22\theta_{13}$, the result in the $\theta_{13}-\delta_{\rm CP}$ plane
   is used to determine the fraction of all CP phases for which there
   is sensitivity at the given confidence level (black arrows). Here we have
taken the $3\sigma$ curve of Fig.~\ref{th13disc1300}.
   Repeating this for all possible values of $\sin^22\theta_{13}$
   yields the figure shown in the middle panel. In the last step one
   chooses a set of values for the CP fraction and translates them
   into values of $\sin^22\theta_{13}$ by using the result from the second 
step.
  \label{cpfexp} }
\end{figure}

\begin{figure}
\centering\leavevmode
\includegraphics[width=0.7\textwidth]{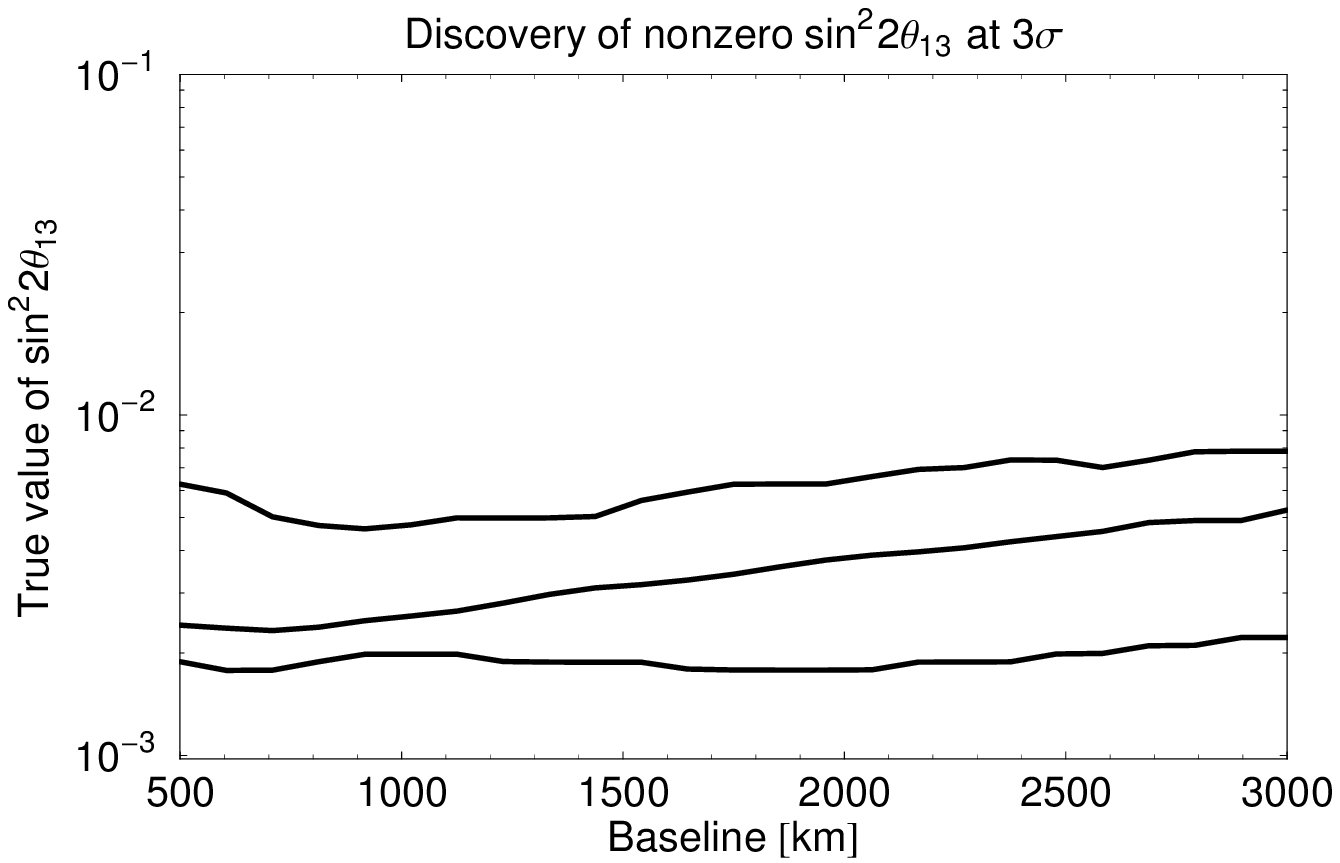}
 \caption{
   Discovery reach for $\sin^22\theta_{13}\neq 0$ at $3\sigma$ for CP
   fractions 0 (lowermost line, best case), 0.5 (middle line) and 1
   (uppermost line, worst case) as a function of the baseline. The
   detector mass, beam power and exposure are kept the same for all baselines.
  \label{th13base} }
\end{figure}

\begin{figure}
\centering\leavevmode
\includegraphics[angle=0,width=0.5\textwidth]{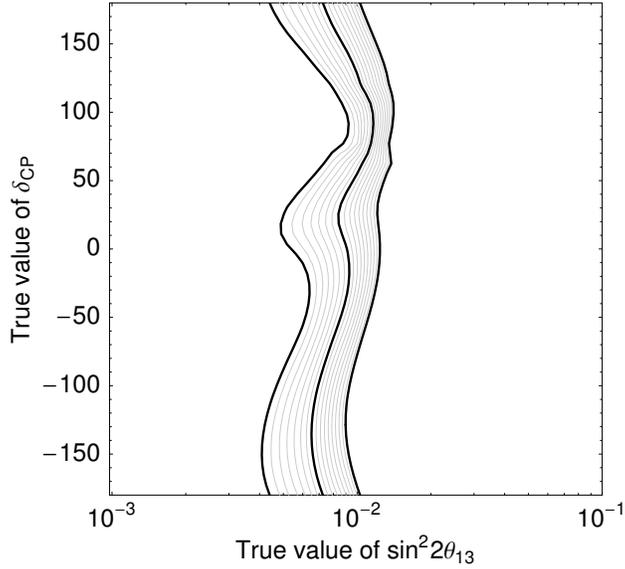}
 \caption{ 
   Discovery potential for a normal mass hierarchy at a baseline of
   $1300\,\mathrm{km}$. 
The bold iso-$\chi^2$ lines are
   $3,4,5\,\sigma$ (from left to right) and the light lines show an
   increase of $\chi^2$ by $1$.
%For all points to the right of the rightmost
%   bold line, the normal hierarchy can be established with at least
%   $5\sigma$ significance. 
For the 
inverted hierarchy the results are approximately 
the same because of the approximately symmetric $\nu$ and $\bar\nu$ running. 
  \label{MH1300} }
\end{figure}

\begin{figure}
\centering\leavevmode
\includegraphics[width=0.7\textwidth]{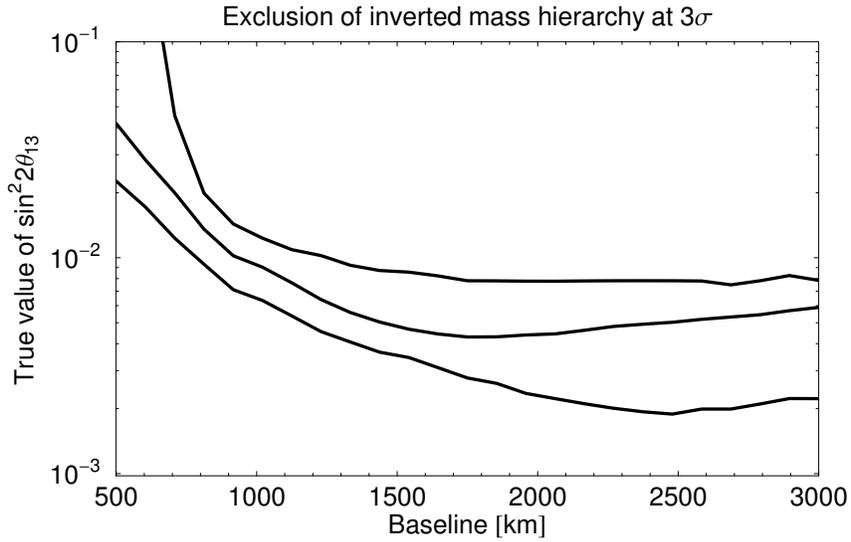}
 \caption{
   Discovery reach for a normal mass hierarchy at $3\sigma$ for CP
   fractions 0 (lowermost line, best case), 0.5 (middle line) and 1
   (uppermost line, worst case) as a function of the baseline. The
   detector mass, beam power and exposure are kept the same  for all 
baselines.
  \label{MHbase} }
\end{figure}

\begin{figure}
\centering\leavevmode
\includegraphics[angle=0,width=0.5\textwidth]{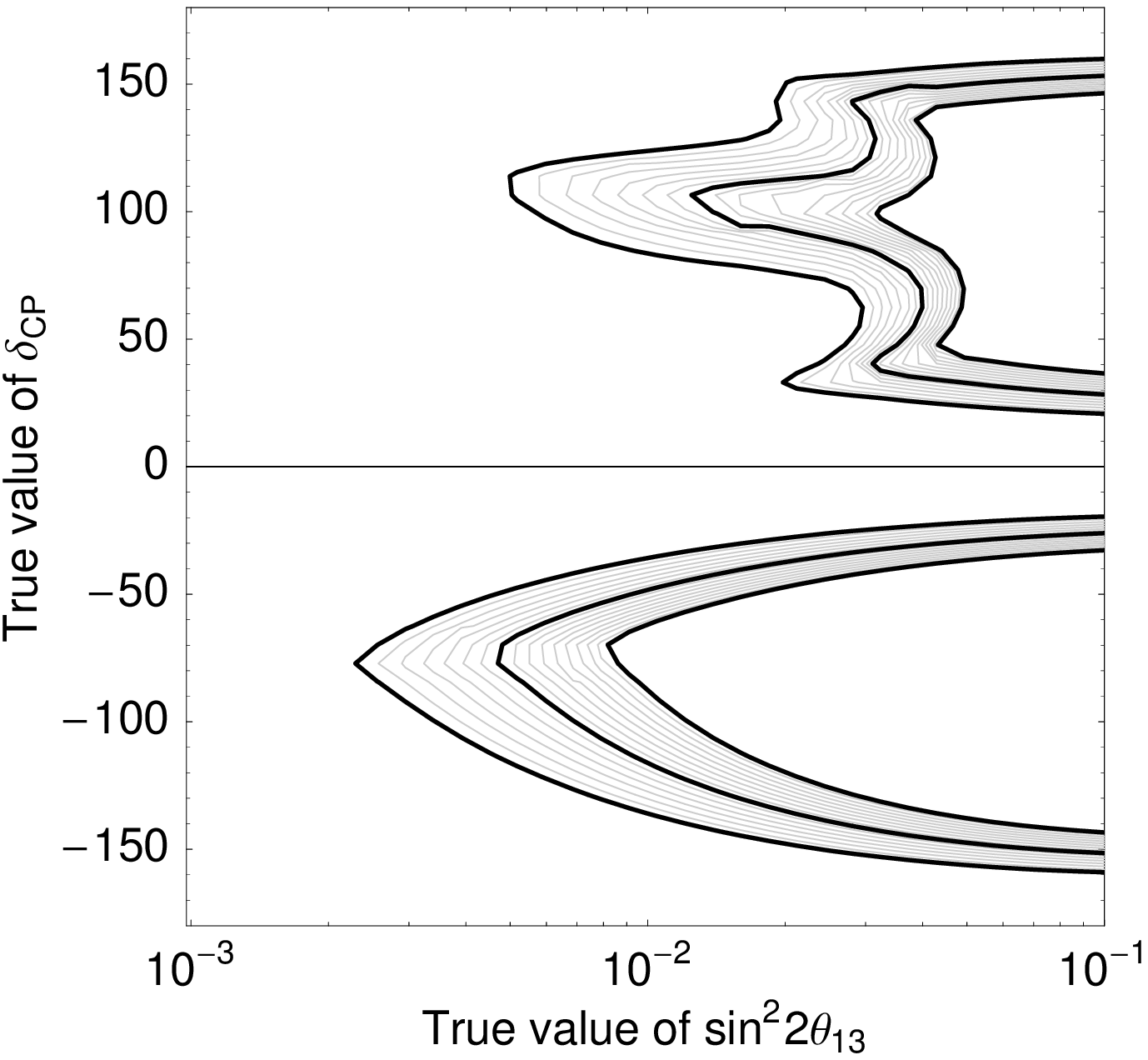}%
\includegraphics[angle=0,width=0.5\textwidth]{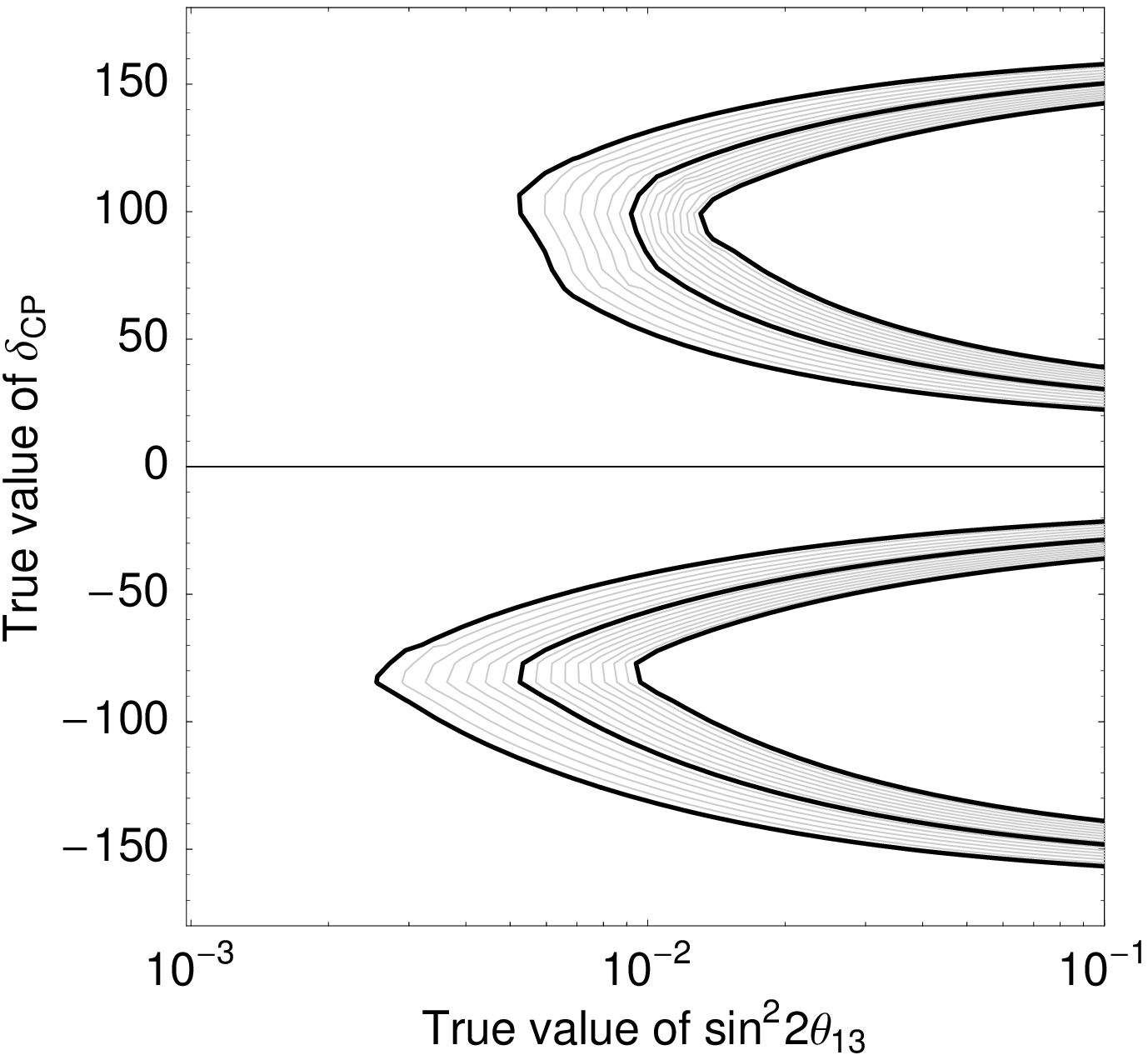}
 \caption{
   Discovery potential for CP violation at baselines of
   $730\,\mathrm{km}$ (left-hand panel) and $1300\,\mathrm{km}$ 
(right-hand panel). 
The bold iso-$\chi^2$ lines are
   $3,4,5\,\sigma$ (from left to right) and the light lines show an
   increase of $\chi^2$ by $1$.
 For all points to the right of the rightmost bold
   line, CP violation can be established with at least $5\sigma$
   significance.
  \label{CPV1300} }
\end{figure}

\begin{figure}
\centering\leavevmode
\includegraphics[width=0.7\textwidth]{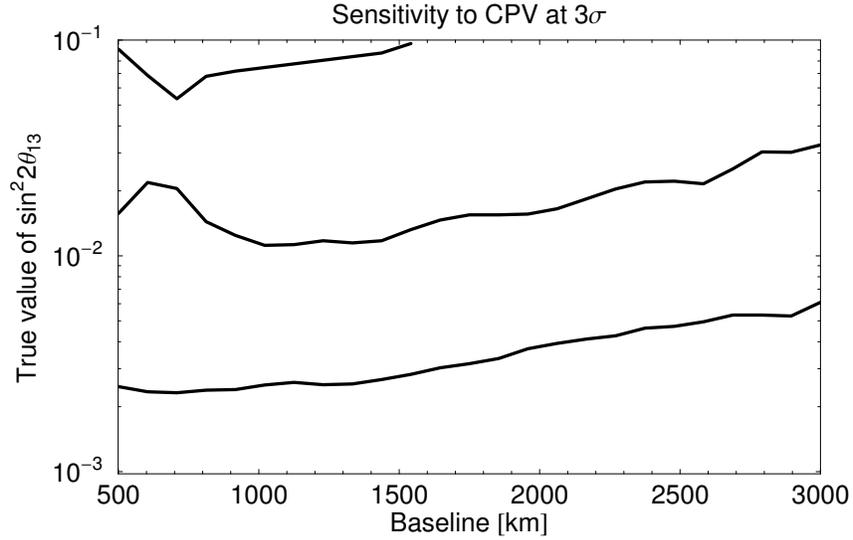}
 \caption{
   Discovery reach for CP violation at $3\sigma$ for CP fractions 0
   (lowermost line, best case), 0.5 (middle line) and 0.75 (uppermost
   line) as a function of the baseline. The detector mass, beam power
   and exposure are kept the same for all baselines.
  \label{CPVbase} }
\end{figure}

\begin{figure}
\centering\leavevmode
  \includegraphics[angle=0,width=0.7\textwidth]{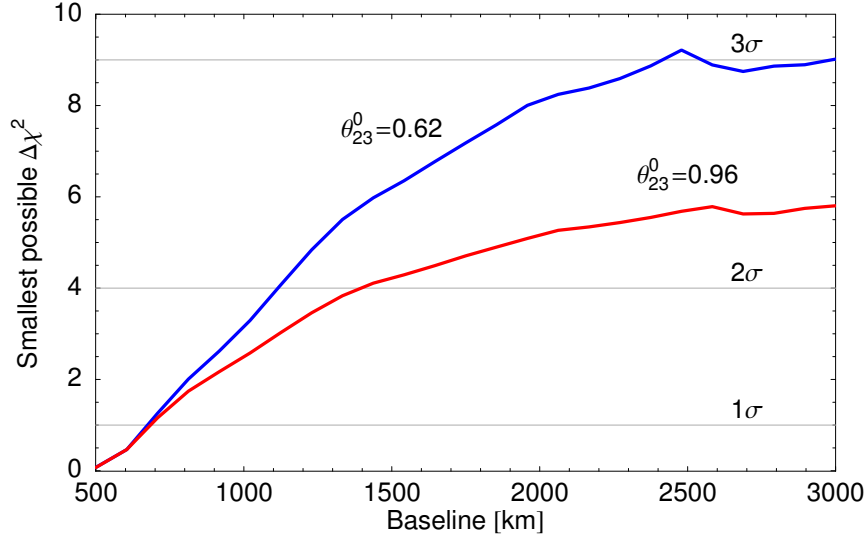}
 \caption{
   Discovery reach for the octant of $\theta_{23}$. Only the most conservative
   case with respect to the true values of $\theta_{13}$ and $\delta_{\rm CP}$
   is considered. The $\chi^2$ difference between
   the true and wrong octant is shown as a function of the baseline for two 
   representative true values of $\theta_{23}$ that are far outside the 
$1\sigma$ range in Eq.~(\ref{equation}) (so as to emphasize how challenging
this measurement is).
The detector mass, beam power and exposure are kept the same for all baselines.
  \label{Cbase} }
\end{figure}

Our results are:

\begin{itemize} 
\item Figure~\ref{th13disc1300} shows the potential for discovering a
  nonzero value of $\theta_{13}$. The sensitivity does not depend
  strongly on the value of $\delta_{\rm CP}$. A finite value of $\theta_{13}$
  can be established at $3\sigma$ for $\sin^2 2\theta_{13}$ as low as
  0.005, irrespective of the value of $\delta_{\rm CP}$.
  
  In what follows, we use the idea of a CP fraction, the definition of
  which is provided in Fig.~\ref{cpfexp} via illustration. For
  example, from the third panel one can see that for 75\% of the
  values of $\delta_{\rm CP}$, the experiment can detect nonzero $\theta_{13}$
  at the $3\sigma$ C.~L. if $\sin^22\theta_{13}\gsim 0.004$.  The most
  optimistic case arises for a CP fraction of 0 since the true value
  of the CP phase is the one for which the sensitivity is maximum.

\item From Fig.~\ref{th13base}, it is evident that the baseline is not
  crucial in determining the ability to confirm if $\theta_{13}$ is
  nonzero. This independence stems from the fact that with increasing
  baseline $L$ the signal event rates do \emph{not} drop as $L^{-2}$
  but much more slowly due to enhancement by matter effects. 
However,
  the background drops  faster than $L^{-2}$ since the
  oscillation with the solar mass splitting reduces the beam intrinsic
  background. This can be seen from Fig.~\ref{rates}. As a result, the
  ratio of signal to the square root of the background is
  approximately constant.
  
\item In Fig.~\ref{MH1300} we address the potential for 
discovering a normal mass
  hierarchy with a baseline of 1300 km. A normal hierarchy can be
  confirmed independently of $\delta_{CP}$ 
  at $>3\sigma$ so long as $\sin^2 2\theta_{13}$ is larger
  than about 0.01. For the inverted hierarchy the result is approximately the 
same because of the nearly symmetric $\nu$ and $\bar\nu$ running.

\item Figure~\ref{MHbase} confirms the expectation that a longer
  baseline improves the sensitivity to the mass hierarchy since the
  matter effects increase with baseline. Note that an experiment with
  baseline below 1000 km has poor sensitivity to the mass hierarchy.
  However, the sensitivity (approximately) plateaus for baselines
  above 1500 km. As we see below, the optimal baseline is determined
  primarily by the discovery reach for CP violation.
 
\item CP violation in the neutrino sector will be discovered if
  $\delta_{\rm CP}=0$ and $\pi$ can be excluded.  Figure~\ref{CPV1300}
  shows the sensitivity for this measurement at two baselines,
  $730\,\mathrm{km}$ (left-hand panel) and $1300\,\mathrm{km}$ 
(right-hand panel).  At a baseline of $730\,\mathrm{km}$ the mass hierarchy
  cannot be resolved and hence the sensitivity for $\delta_{\rm CP}>0$ (for
  the inverted hierarchy, this occurs for $\delta_{\rm CP}<0$) is
  severely limited by so-called $\pi$-transit: for certain
  combinations of true $\theta_{13}$ and $\delta_{\rm CP}$ it is possible to
  fit the data with the wrong mass hierarchy and
  $\delta_{\rm CP}=\pi$~\cite{Huber:2002mx}. The existence of this solution
  inhibits the ability to discover CP violation for a range of 
  $\delta_{\rm CP}$ in an experiment with 
  baseline from FNAL to the Soudan mine. At a baseline of
  $1300\,\mathrm{km}$ the mass hierarchy is well resolved (see
  Fig.~\ref{MH1300}) and therefore no problems arise due to $\pi$-transit.  
  Baselines below
  approximately $1000\,\mathrm{km}$ are not optimal for studying CP
  violation. However, also at longer baselines it can be seen that
  the sensitivity to CP violation is greater for $\delta_{\rm CP}<0$.
  This is mainly due to the somewhat lower statistical significance of
  the antineutrino data; in the antineutrino mode, the signal event
  rate is lower and the background event rate is higher.

\item In Fig.~\ref{CPVbase}, we show results for a CP fraction equal
  to 0.75 instead of for unity, since it is impossible for an
  experiment to have sensitivity to CP violation if $\delta_{\rm CP}=0$ or
  $\pi$. Note that the sensitivity  decreases with baseline
  above 1000 km, which is to be attributed to the competition of
  genuine CP violation and fake CP violation induced by matter effects. 
Nevertheless, the dependence of the sensitivity on the solar and atmospheric
oscillation parameters  and on
 the  length of the baseline beyond $1000$ km
is weak and could be partly compensated by optimizing the
low energy part of the spectrum. The flexibility at FNAL to vary the proton 
energy while keeping the total beam power constant  is important for this 
optimization.      
%It is well known that at very long
%  baseline, CP effects are effectively suppressed by the presence
%  of the matter potential.  
Experiments with beams from FNAL to
  the Homestake mine or the Henderson mine, which have baselines of
  1300 km and 1500 km, respectively, have good sensitivity to
  CP violation compared to the optimal case of 1000 km and also
  have very good sensitivity to $\theta_{13}$ and the
  mass hierarchy. Thus, baselines around \mbox{1300 -- 1500 km} represent a
  global optimum.

\item We now consider the prospect of breaking the $(\theta_{23},
  \pi/2-\theta_{23})$ degeneracy, which is only present if the true
  value of $\theta_{23}\neq\pi/4$. We display the results for two
  possible true values of $\theta_{23}$ in Fig.~\ref{Cbase}. To emphasize
how challenging it is to break this degeneracy, we have selected values 
of $\theta_{23}$ that are far outside the $1\sigma$ 
 range in Eq.~(\ref{equation}). The
  ability to exclude the wrong octant depends on the choice of the
  true values of $\theta_{13}$ and $\delta_{\rm CP}$. This dependence is not
  pronounced for most parts of the parameter space and therefore we
  chose to consider only the most pessimistic case, {\it i.e.} only that
  combination of true $\theta_{13}$ and $\delta_{\rm CP}$ which gives the 
  lowest  $\chi^2$ difference between the true and wrong octant. Hence, the
  results shown in Fig.~\ref{Cbase} are conservative for the representative
true values of $\theta_{23}$. Clearly this
  measurement favors baselines longer than 2000~km.  As noted earlier, such
  long baselines are not preferred for the detection of CP violation.
  Distinguishing the octant is a very difficult
  measurement which cannot be guaranteed to succeed with high
  confidence. If the neutrino spectrum at low energies can be enhanced,
perhaps in a separate  run with lower energy protons or off-axis 
geometry,  then the signal to background ratio for this measurement 
could be improved.  
%We emphasize that 
%the flexibility at FNAL to provide 
%protons of varying energy at constant power is important for 
%this physics. 

\item Finally, we assess how our results depend on exposure and on the
  uncertainty in the overall normalization of the background. 
  We have  not considered the 
 systematic uncertainties in the
  shape of the background assuming that 
  the background shape,  which is sharply
  peaked at low energies,  will be measured sufficiently well 
by a near detector, as in the MINOS experiment~\cite{minos}. 
  
  From Fig.~\ref{lumi} we see that reducing the exposure by a factor
  of 3, degrades the sensitivity to a nonzero $\sin^2 2\theta_{13}$
  and to determine the mass hierarchy by about a factor of 2 in
  $\sin^2 2\theta_{13}$, while the sensitivity to CP violation worsens
  by a factor of about $2-5$ depending on the value of $\delta_{\rm CP}$; if
  $\delta_{\rm CP}$ has an unfavorable value, the experiment needs the
  highest possible exposure to maintain its ability to detect CP
  violation. Nevertheless, operation with half the total beam power, 
 which could be achieved by  the accumulator based  upgrades at FNAL,  
  is shown to have excellent sensitivity.

  Similar statements can be made for the dependence on the systematic
  uncertainty from Fig.~\ref{sys}. Again, the sensitivity to CP
  violation puts the most stringent requirements on the experiment.
  The most striking conclusion is that sensitivity to the mass
  hierarchy is only weakly affected by the systematic uncertainty.  

\end{itemize}

\section{Summary}

We have studied  the measurement of CP violation in 
the neutrino sector  using a powerful, but conventional 
neutrino beam, using a MW-class proton source located in the U.S.
(either at 
FNAL or BNL). The beam would be delivered 
 to a large detector with fiducial mass $\sim 300$kT 
over a distance $\gsim 1000$km. Such a detector could be 
built at a new Deep Underground Science and Engineering Laboratory 
(DUSEL) which is in the process of being evaluated in the U.S.  
The two possible sites for DUSEL at the Homestake mine in South Dakota and 
the Henderson mine in Colorado are at distances of 1290 km and 1487 km
from FNAL, respectively (or 2540 km and 2770 km from BNL, respectively). 
We have calculated the scientific reach and its 
dependence on total exposure and systematics for a range of distances
that encompass these possibilities.  

%It is important to
%recognize that the detector designed for such an experiment needs to be
%highly capable in terms of pattern recognition and energy
%resolution. If such a detector is located in a deep low background
%environment, it has broad applications in searching for nucleon decay
%and astrophysical neutrino sources.  

The experiment is motivated by
the need to have sensitivity to both the
atmospheric and solar oscillation scales
(a necessary condition for observing CP 
violation in the 3-generation framework), and to obtain
an oscillatory pattern in the energy spectrum of  neutrinos.
Access to multiple nodes of oscillations leads to improved sensitivity 
for CP violation because the CP effect grows larger 
for successive nodes and because parameter ambiguities can be 
resolved.    

 We have  shown that the 
sensitivity to nonzero $\theta_{13}$ 
using the appearance of electron neutrinos and antineutrinos 
is roughly independent of baseline in the range of 1000 to 3000~km.
Also, the sensitivity to the mass hierarchy plateaus for baselines 
above 1500 km.  
On the other hand, sensitivity to CP violation decreases with
baseline above 1000 km, but it is a weak function of the baseline 
beyond 1000 km. 
 The baselines for a beam from FNAL to Homestake
or Henderson are optimal for establishing a nonzero $\theta_{13}$, the
mass hierarchy and CP violation.
The determination of the $\theta_{23}$ octant, however, favors baselines longer than 1500 km. 
Both the  $\theta_{23}$  measurement and the CP measurement could be improved by
increasing  the low energy part of the flux. 
%Unless $\theta_{13}$ is close
%to its current upper bound, it is unlikely that the octant of $\theta_{23}$
%will be known from such an experiment.
The sensitivities for a 1300 km baseline are shown in 
Figs.~\ref{th13disc1300}, \ref{MH1300} and \ref{CPV1300}. For the same
baseline, the dependence of the sensitivities on exposure and systematic
uncertainties are summarized in 
Figs.~\ref{lumi} and \ref{sys}, respectively.

\section{Acknowledgments}
This research was supported by the DOE
under Grants No.~DE-FG02-95ER40896, DE-AC02-98CH10886 and DE-FG02-04ER41308, 
by the NSF under CAREER Award No.~PHY-0544278 and Grant No.~EPS-0236913, 
by the State of Kansas through the
Kansas Technology Enterprise Corporation, and by the KU General Research
Fund Program.
Computations were performed on facilities
supported by the
NSF under Grants No. EIA-032078 (GLOW), PHY-0516857
(CMS Research Program subcontract from UCLA), and PHY-0533280
(DISUN), and by the University of Wisconsin Graduate
School/Wisconsin Alumni Research Foundation.

%\newpage
%\input{biblio.tex}
%

\newpage

\begin{figure}
\centering\leavevmode
\includegraphics[angle=0,width=1\textwidth]{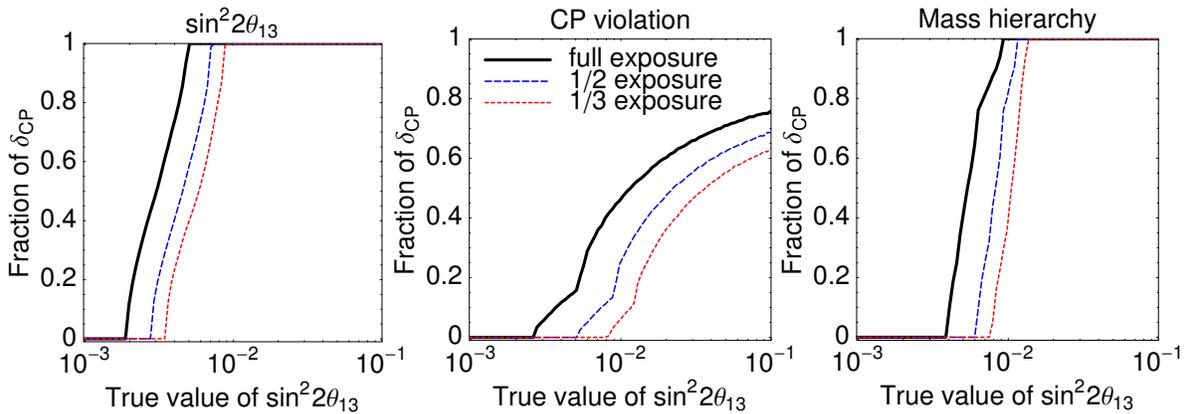}
 \caption{Dependence of the $3\sigma$ sensitivities on exposure for an
   experiment with 1300~km baseline. The two cases considered are
   $1/2$ and $1/3$ of the (full)  exposure we have used
   throughout.
   \label{lumi} }
\end{figure}

\begin{figure}
\centering\leavevmode
\includegraphics[angle=0,width=1\textwidth]{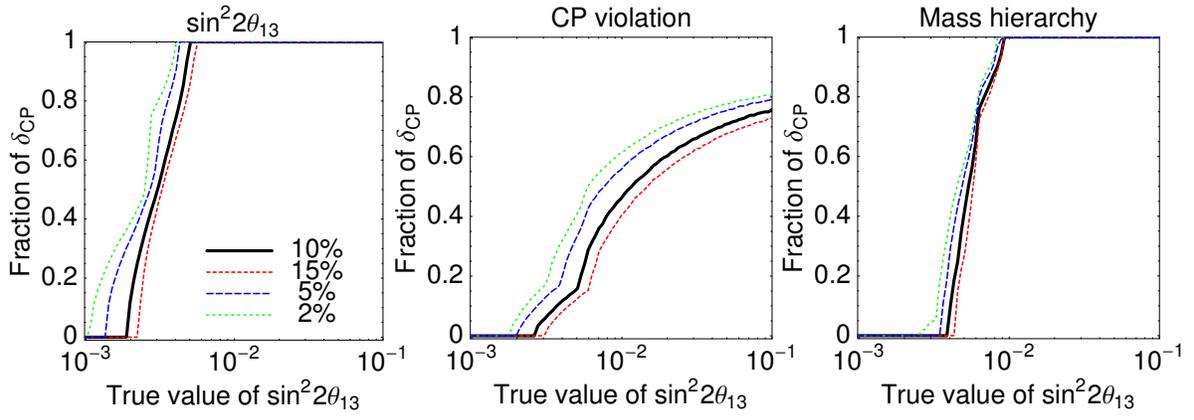}
 \caption{Dependence of the $3\sigma$ sensitivities on the uncertainty in the 
   overall normalization of the background for an experiment with a
   1300~km baseline. We have adopted a 10\% uncertainty throughout.
  \label{sys} }
\end{figure}

\end{document}